\pdfoutput=1
\documentclass[acmsmall]{acmart}

\usepackage[T1]{fontenc}
\usepackage[utf8]{inputenc}
\usepackage[english]{babel}

\renewcommand\footnotetextcopyrightpermission[1]{} 

\usepackage{mathtools,amsfonts}
\usepackage{nicefrac}
\usepackage{graphicx}
\usepackage{xcolor}
\usepackage{listings}
\usepackage{csquotes}
\usepackage{tikz}
\usepackage[binary-units=true]{siunitx}
\usepackage[textsize=scriptsize]{todonotes}
\usepackage{subcaption}
\usepackage{hyperref}

\usetikzlibrary{arrows.meta,calc,positioning,matrix,shapes.geometric}
\usetikzlibrary{quantikz}

\acmYear{202z}
\acmJournal{TQC}
\acmVolume{x}
\acmNumber{y}
\acmArticle{}
\acmMonth{0}

\tikzset{%
    terminal/.style={draw, rectangle, inner sep=0pt, minimum width=0.5cm, minimum height=0.5cm},
    vertex/.style={draw, circle, inner sep=0pt,minimum width=0.5cm,minimum height=0.5cm}
}

\newtheorem{example}{Example}
\newcommand{\qop}[1]{\ensuremath{\mathit{#1}}}

\begin{document}

\title{Tools for Quantum Computing Based on Decision Diagrams}
\author{Robert Wille}
\affiliation{%
   \institution{Johannes Kepler University Linz}
   \city{Linz}
   \postcode{4040}
   \country{Austria}
}
\email{robert.wille@jku.at}
\orcid{0000-0002-4993-7860}
\affiliation{%
  \institution{Software Competence Center Hagenberg GmbH (SCCH)}
  \city{Hagenberg}
  \postcode{4232}
  \country{Austria}
}
\author{Stefan Hillmich}
\affiliation{%
   \institution{Johannes Kepler University Linz}
   \city{Linz}
   \postcode{4040}
   \country{Austria}
}
\email{stefan.hillmich@jku.at}
\orcid{0000-0003-1089-3263}
\author{Lukas Burgholzer}
\affiliation{%
   \institution{Johannes Kepler University Linz}
   \city{Linz}
   \postcode{4040}
   \country{Austria}
}
\email{lukas.burgholzer@jku.at}
\orcid{0000-0003-4699-1316}

\begin{abstract}
With quantum computers promising advantages even in the near-term NISQ era, there is a lively community that develops software and toolkits for the design of corresponding quantum circuits. 
Although the underlying problems are different, expertise from the design automation community, which developed sophisticated design solutions for the conventional realm  in the past decades, can help here.
In this respect, decision diagrams provide a promising foundation for tackling many design tasks such as simulation, synthesis, and verification of quantum circuits. 
However, users of the corresponding tools often do not have a proper background or an intuition about how these methods based on decision diagrams work and what their strengths and limits are.
In this work, we first review the concepts 
of how decision diagrams can be employed, e.g., for  
the simulation and verification of quantum circuits.
Afterwards, in an effort to make decision diagrams for quantum computing more accessible, we then present a visualization tool for quantum decision diagrams, which allows users to explore the behavior of decision diagrams in the design tasks mentioned above. 
Finally, we present decision diagram-based tools for simulation and verification of quantum circuits using the methods discussed above as part of the open-source JKQ quantum toolset---a set of tools for quantum computing developed at the \emph{Johannes Kepler University (JKU) Linz} and released under the MIT license.
More information about the corresponding tools is available at \url{https://github.com/iic-jku/}.
By this, we provide an introduction of the concepts and tools for potential users who would like to work with them as well as potential developers aiming to extend them.
\end{abstract}


\maketitle
\section{Introduction}

Quantum computers are steadily improving in terms of their computational power to the extent that first computations are being performed that are no longer feasible on conventional machines~\cite{aruteQuantumSupremacyUsing2019,zhongQuantumComputationalAdvantage2020}. 
Achieving these milestones is only possible through interdisciplinary efforts by physicists, mathematicians, computer scientists, and many others. 
Just as in the design of conventional circuits and systems, the development of design automation tools for quantum computing will be one of the key factors for the success of the technology.

In the 80's, decision diagrams were proposed as a data structure for efficient representation and manipulation of Boolean functions~\cite{bryantGraphbasedAlgorithmsBoolean1986}.
Following this development, a multitude of decision diagrams such as BDDs, FBDDs, KFDDs, MTBDDs, and ZDDs emerged (see, e.g.,~\cite{bryantSymbolicBooleanManipulation1992,wegenerBranchingProgramsBinary2000,gergovEfficientBooleanManipulation1994,drechslerEfficientRepresentationManipulation1994,baharAlgebraicDecisionDiagrams1993,minatoZerosuppressedBDDsSet1993}), which established them as a core asset in the design of today's circuits and systems. 
Due to their success in the past, the use of decision diagrams has also been proposed in the domain of quantum computing~\cite{viamontesImprovingGatelevelSimulation2003,millerQMDDDecisionDiagram2006,wangXQDDbasedVerificationMethod2008,niemannQMDDsEfficientQuantum2016,DBLP:conf/iccad/ZulehnerHW19,hongTensorNetworkBased2020,vinkhuijzenLIMDDDecisionDiagram2021}.
In particular for design tasks such as \emph{simulation}~\cite{viamontesImprovingGatelevelSimulation2003,DBLP:journals/tcad/ZulehnerW19,hillmich2020just,vinkhuijzenLIMDDDecisionDiagram2021}, \emph{synthesis}~\cite{niemannEfficientSynthesisQuantum2014,abdollahiAnalysisSynthesisQuantum2006,soekenSynthesisReversibleCircuits2012,DBLP:journals/tcad/ZulehnerW18}, and \emph{verification}~\cite{wangXQDDbasedVerificationMethod2008,burgholzerAdvancedEquivalenceChecking2021,smithQuantumLogicSynthesis2019,hongEquivalenceCheckingDynamic2021} of quantum circuits, they found great interest recently.

Despite the vast knowledge on design automation that may be exploited in quantum computing, there is still a huge gap to bridge between the quantum computing community and the design automation community.
In fact, many promising techniques have hardly reached the core of the quantum computing community until now and are not yet established---despite showing promising results.
Due to the interdisciplinarity of the field, quite often, users of the corresponding tools are hardly familiar with the underlying concepts and, understandably, have a hard time getting 
a proper intuition about how these tools work.

With this work, we aim at narrowing the aforementioned gap. 
We show that decision diagrams are a promising tool for design automation in the quantum realm.
After defining and reviewing decision diagrams for quantum computing, we show how an easy-to-use visualization can help to develop an intuition on the behavior of decision diagrams.
The software we present focuses on simulation as well as verification of quantum circuits and offers different styles to accommodate various use cases.
Subsequently, we introduce our decision diagram-based tools for simulation and verification as part of the open-source JKQ quantum toolset---a set of corresponding tools developed at the \emph{Johannes Kepler University (JKU) Linz}. 
While each tool (including ours) certainly has strengths and weaknesses, we offer complementary approaches for many of the problems that need to be tackled when designing quantum circuits.
By making our tools publicly available as open-source, we also provide other researchers with the option to incorporate the underlying methods into their existing tools. In fact, this already motivated ``players'' like IBM and Atos to include, e.g., the simulation approach based on decision diagrams into their tools.

The remainder of this work is structured as follows.
In \autoref{sec:ddsapp}, we review decision diagrams and their applications in quantum computing.
\autoref{sec:visualization} illustrates how the application of decision diagrams in simulation and verification of quantum circuits can be visualized to help the interested user to develop an intuition.
In \autoref{sec:user}, we detail how to approach simulation and verification with the \emph{JKQ} tools from a user's perspective.
\autoref{sec:developer} gives insight on how the \emph{JKQ} tool set is organized and where interested developers can start to extend the available tools with their own methods.
Finally, we conclude this work in \autoref{sec:conclusions}.

\section{Decision Diagrams and their Applications}\label{sec:ddsapp}

State vectors and operation matrices of a quantum system are exponential in size with respect to the number of qubits---quickly rendering the representation of a system state or the construction of a system matrix an extremely difficult task.
\emph{Decision diagrams} have been proposed as an efficient way for representing and manipulating quantum functionality~\cite{niemannQMDDsEfficientQuantum2016,DBLP:conf/iccad/ZulehnerHW19,millerQMDDDecisionDiagram2006,viamontesImprovingGatelevelSimulation2003,DBLP:journals/tcad/ZulehnerW19,hillmich2020just,niemannEfficientSynthesisQuantum2014,abdollahiAnalysisSynthesisQuantum2006,soekenSynthesisReversibleCircuits2012,wangXQDDbasedVerificationMethod2008,burgholzerAdvancedEquivalenceChecking2021,smithQuantumLogicSynthesis2019,DBLP:journals/tcad/ZulehnerW18, hongTensorNetworkBased2020,vinkhuijzenLIMDDDecisionDiagram2021, hongEquivalenceCheckingDynamic2021}.
While they are still exponential in the worst-case, decision diagrams have been shown to lead to very compact representations in many cases.
In the following, we review how decision diagrams for quantum computing work and how they can be applied to the problems of \emph{quantum circuit simulation} and \emph{quantum circuit verification}.

\subsection{Decision Diagrams}\label{sec:dds}

\begin{figure}[tbp]
    \centering
    \begin{gather*}
        [\alpha]^\top = [\alpha_{00\ldots 0}, \hdots, \alpha_{11\ldots 1}]^\top \\
        	[\alpha_{\boldsymbol{0}0\ldots0},\ldots,\alpha_{\boldsymbol{0}1\ldots1}]^\top \qquad\qquad\qquad\qquad
        	[\alpha_{\boldsymbol{1}0\ldots0},\ldots,\alpha_{\boldsymbol{1}1\ldots1}]^\top \\
        		[\alpha_{\boldsymbol{00}0\ldots0},\ldots,\alpha_{\boldsymbol{00}1\ldots1}]^\top \quad
        		[\alpha_{\boldsymbol{01}0\ldots0},\ldots,\alpha_{\boldsymbol{01}1\ldots1}]^\top \qquad 
        		[\alpha_{\boldsymbol{10}0\ldots0},\ldots,\alpha_{\boldsymbol{10}1\ldots1}]^\top \quad 
        		[\alpha_{\boldsymbol{11}0\ldots0},\ldots,\alpha_{\boldsymbol{11}1\ldots1}]^\top\\
        \vdots
    \end{gather*}
    \caption{Decomposition of a state vector}
    \label{fig:decomposition}
\end{figure}

The state of an $n$-qubit system is represented by a state vector of size $2^n$---an exponential representation.
However, the inherent tensor product structure of many quantum states and redundancies in their description provide ground for a more compact representation.
To this end, a given state vector \(\alpha\) with its complex amplitudes \(\alpha_i\) for \(i \in \{0,1\}^n\) is decomposed into sub-vectors as illustrated in \autoref{fig:decomposition} until only individual complex numbers remain.

This gives rise to a decision diagram structure with $n$ levels of nodes (labeled $q_{n-1}$ to $q_0$) and the individual amplitudes as its leaves.
Each node \(q_i\) has exactly two successors---indicating whether the corresponding path leads to an amplitude where qubit $q_i$ is in the state \ket{0} or \ket{1}, respectively.
During these decompositions, equivalent \mbox{sub-vectors} can be represented by the same node---allowing for sharing and, hence, a reduction of the complexity of the representation.
Further compaction is achieved by introducing edge weights on all levels and employing normalization schemes\footnote{Normalization can be performed by, e.g., dividing the weight of the outgoing edges of a node by the norm of the vector containing both edge weights and multiplying this factor to the incoming edges---allowing for efficient sampling from the resulting decision diagram~\cite{hillmich2020just}.}, in order to unify \mbox{sub-vectors} only differing by a common factor and further exploit possible redundancies.
The amplitude of a given basis state can then be reconstructed from the multiplication of the edge weights along the path from the root node to the terminal node.
An example illustrates the idea.

\begin{figure}[tp]
	\centering
	\begin{subfigure}[b]{0.37\linewidth}
	\centering
	\begin{tikzpicture}[terminal/.style={draw,rectangle,inner sep=0pt}]	
			\matrix[ampersand replacement=\&,every node/.style={draw,circle,inner sep=0pt,minimum width=0.5cm,minimum height=0.5cm},column sep={0.5cm,between origins},row sep={1cm,between origins}] (qmdd) {
								\& \node (m1) {$q_1$};           \& \\
				\node (m0a) {$q_0$};   \&  				     \& \node (m0b) {$q_0$}; \\
				                \& \node[terminal] (t3) {$1$}; \& \\
			};
			
			\draw[] ($(m1)+(0,0.7cm)$) -- (m1);
			
			\draw[] (m1) -- ++(240:0.6cm) node[left, near start] {$\frac{1}{\sqrt{2}}$}-- (m0a);
			\draw (m1) -- ++(300:0.6cm) node[right, near start] {$\frac{1}{\sqrt{2}}$} -- (m0b);
			\draw (m0a) -- ++(240:0.6cm) -- (t3);
			\draw (m0a) -- ++(300:0.4cm) node[below, xshift=0.5pt, inner sep=0,font=\scriptsize] {$0$};			
			\draw (m0b) -- ++(240:0.4cm) node[below, xshift=-0.5pt, inner sep=0,font=\scriptsize] {$0$};
			\draw (m0b) -- ++(300:0.6cm) -- (t3);
		\end{tikzpicture}
	\caption{$\ket{\varphi}=\nicefrac{1}{\sqrt{2}}\,[1,\,0,\,0,\,1]^\top$}
	\label{fig:belldd}
	\end{subfigure}%
	\hfill
	\begin{subfigure}[b]{0.3\linewidth}
	\centering
	\begin{tikzpicture}
		\node[draw,circle,inner sep=0pt,minimum width=0.5cm,minimum height=0.5cm] (n1) {$q_{0}$};
		\node[below = 0.45cm of n1, terminal] (t) {$1$};
		\draw (n1) -- ++(220:0.6cm) -- (t);
		\draw (n1) -- ++(240:0.5cm) -- (t);
		\draw (n1) -- ++(300:0.5cm) -- (t);
		\draw (n1) -- ++(320:0.6cm) node[right, near start] {$-1$} -- (t);
		\draw ($(n1)+(0,0.6cm)$) -- node[right, near start] {$\frac{1}{\sqrt{2}}$} (n1);
	\end{tikzpicture}
	\caption{Hadamard gate}\label{fig:hdd}
	\end{subfigure}%
	\hfill
	\begin{subfigure}[b]{0.32\linewidth}
	\centering
	\begin{tikzpicture}
		\matrix[ampersand replacement=\&, column sep={0.5cm,between origins}, row sep={1cm,between origins}] (qmdd) {
								\& \node[vertex] (m1) {$q_1$};           \& \\
				\node[vertex] (m0a) {$q_0$};   \&  				     \& \node[vertex] (m0b) {$q_0$}; \\
				                \& \node[terminal] (t3) {$1$}; \& \\
			};
			
			\draw[] ($(m1)+(0,0.7cm)$) -- (m1);
			
			\draw (m1) -- ++(220:0.6cm) -- (m0a);
			\draw (m1) -- ++(240:0.4cm) node[below, xshift=-0.5pt, inner sep=0,font=\scriptsize] {$0$};			
			\draw (m1) -- ++(300:0.4cm) node[below, xshift=0.5pt, inner sep=0,font=\scriptsize] {$0$};			
			\draw (m1) -- ++(320:0.6cm) -- (m0b);
			\draw (m0a) -- ++(220:0.7cm)  -- (t3);
			\draw (m0a) -- ++(240:0.4cm) node[below, xshift=-0.5pt, inner sep=0,font=\scriptsize] {$0$};			
			\draw (m0a) -- ++(300:0.4cm) node[below, xshift=0.5pt, inner sep=0,font=\scriptsize] {$0$};
			\draw (m0a) -- ++(320:0.6cm)  -- (t3);

			\draw (m0b) -- ++(220:0.4cm) node[below, xshift=-0.5pt, inner sep=0,font=\scriptsize] {$0$};
			\draw (m0b) -- ++(250:0.4cm) -- (t3);
			\draw (m0b) -- ++(290:0.4cm) -- (t3);			
			\draw (m0b) -- ++(320:0.4cm) node[below, xshift=-0.5pt, inner sep=0,font=\scriptsize] {$0$};
	\end{tikzpicture}
	\caption{Controlled-NOT gate}\label{fig:cnotdd}
	\end{subfigure}
	\caption{Decision diagram representations for quantum states and operations}
\end{figure}
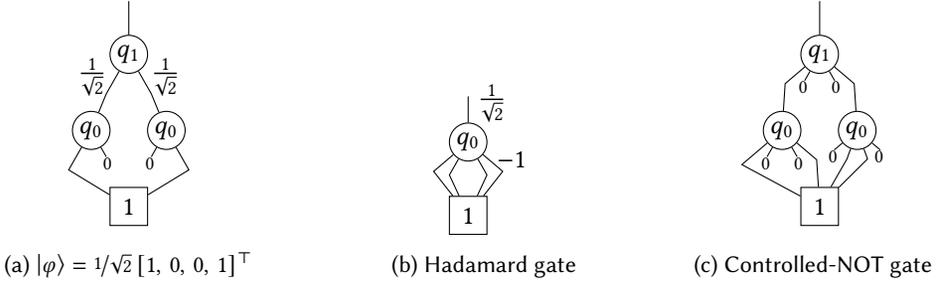

\begin{example}\label{ex:dd}
	Consider the state $\ket{\varphi} \equiv \nicefrac{1}{\sqrt{2}}\,[1,\,0,\,0,\,1]^\top$. 
	Recursively splitting this vector into sub-vectors results in a decision diagram as shown in \autoref{fig:belldd}. 
	It consists of 3 nodes (the terminal node is usually not counted towards a decision diagram's size).
	The two paths leading from the root edge to the terminal node encode the states $\ket{00}$ and $\ket{11}$ respectively---each with an amplitude of \mbox{$\nicefrac{1}{\sqrt{2}}\,\cdot\, 1 = \nicefrac{1}{\sqrt{2}}$}.
	Sub-vectors composed solely of $0$ entries are typically denoted by $0$-stubs to reduce visual clutter, while edge weights equal to $1$ are frequently omitted.
\end{example}

A similar construction is employed for representing quantum operations, i.e., the corresponding unitary matrices. 
Instead of two successors in the decomposition of vectors, each node in a decision diagram representing the (unitary) matrix $U$ of an operation has four successors---corresponding to four equally sized sub-matrices $U_{ij}$ as in
\begin{equation*}
	U = \begin{bmatrix}
    U_{00} & U_{01} \\ U_{10} & U_{11}
    \end{bmatrix}.
\end{equation*}
At each level $l$, this splitting corresponds to the action of~$U$ depending on the value of the qubit $q_l$, i.e.,~ $U_{ij}$ describes how the rest of the system is transformed given that~$q_{l}$ is mapped from~$\ket{j}$ to~$\ket{i}$ for $i,j\in\{0,1\}$.

\begin{example}\label{ex:opdds}
	Consider the single-qubit Hadamard operation and the two-qubit controlled-NOT operation, i.e.,
	\[ 
	    \qop{H} = \frac{1}{\sqrt{2}}\begin{bmatrix*}[r]1 & 1 \\ 1& -1\end{bmatrix*} \qquad \mbox{and} \qquad
	    \qop{CNOT} = \begin{bsmallmatrix}1 & 0 & 0 & 0 \\ 0 & 1 & 0 & 0 \\ 0 & 0 & 0 & 1 \\ 0 & 0 & 1 & 0\end{bsmallmatrix}\mbox{, respectively.}
	\]
	Their corresponding representations as decision diagrams are shown in \autoref{fig:hdd} and \autoref{fig:cnotdd}, respectively. 
	To this end, the first (second) edge points to the node corresponding to the upper-left (upper-right) sub-matrix $U_{00}$ ($U_{01}$),
	while the third (fourth) edge points to the lower-left (lower-right) \mbox{sub-matrix} $U_{10}$ ($U_{11}$).
\end{example}

Matrices of individual gates have to be extended to the full system size using tensor products before being applied to the current state of a system. This extension is a natural operation on decision diagrams.
Given two decision diagrams representing matrices $U$ and $V$, their tensor product $U\otimes V$ is obtained by just replacing the terminal node in the decision diagram of $U$ with the root node of $V$'s decision diagram (and potentially relabeling the nodes).

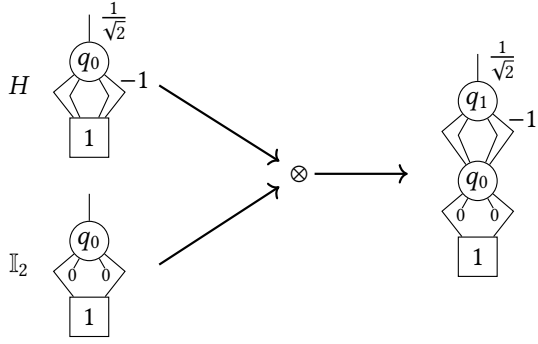
\begin{figure}[tb]
\centering
\resizebox{0.53\linewidth}{!}{
\begin{tikzpicture}
	\node[draw,circle,inner sep=0pt,minimum width=0.5cm,minimum height=0.5cm] (n1) {$q_{0}$};
	\node[below = 0.45cm of n1, terminal] (t1) {$1$};
	\draw (n1) -- ++(220:0.6cm) -- (t1);
	\draw (n1) -- ++(240:0.5cm) -- (t1);
	\draw (n1) -- ++(300:0.5cm) -- (t1);
	\draw (n1) -- ++(320:0.6cm) node[right, near start] {$-1$} -- (t1);
	\draw ($(n1)+(0,0.6cm)$) -- node[right, near start] {$\frac{1}{\sqrt{2}}$} (n1);
				
	\node[draw,circle,inner sep=0pt,minimum width=0.5cm,minimum height=0.5cm, below= 0.8 of t1] (n2) {$q_{0}$};
	\node[below = 0.45cm of n2, terminal] (t2) {$1$};
	\draw (n2) -- ++(220:0.6cm) -- (t2);
	\draw (n2) -- ++(240:0.4cm) node[below, xshift=-0.5pt, inner sep=0,font=\scriptsize] {$0$};
	\draw (n2) -- ++(300:0.4cm) node[below, xshift=0.5pt, inner sep=0,font=\scriptsize] {$0$};
	\draw (n2) -- ++(320:0.6cm) -- (t2);
	\draw ($(n2)+(0,0.6cm)$) -- (n2);
		
	\node[thick] (ot) at ($(t1)!0.35!(n2) + (2.7, 0.0)$) {$\otimes$};
	\draw[->, thick] ($(t1)!0.35!(n2) + (2.9, 0)$) --++(1.2, 0.0);
	\draw[->, thick] ($(n1)+(0.9, -0.3)$) -- (ot);
	\draw[->, thick] ($(n2)+(0.9, -0.3)$) -- (ot);
	
	\node[draw,circle,inner sep=0pt,minimum width=0.5cm,minimum height=0.5cm] (n3) at ($(n1)+(5.0, -0.5)$) {$q_{1}$};
	\node[draw,circle,inner sep=0pt,minimum width=0.5cm,minimum height=0.5cm, below= 0.5 of n3] (n4) {$q_{0}$};

	\draw (n3) -- ++(220:0.6cm) -- (n4);
	\draw (n3) -- ++(240:0.5cm) -- (n4);
	\draw (n3) -- ++(300:0.5cm) -- (n4);
	\draw (n3) -- ++(320:0.6cm) node[right, near start] {$-1$} -- (n4);
	\draw ($(n3)+(0,0.6cm)$) -- node[right, near start] {$\frac{1}{\sqrt{2}}$} (n3);
				
	\node[below = 0.45cm of n4, terminal] (t3) {$1$};
	\draw (n4) -- ++(220:0.6cm) -- (t3);
	\draw (n4) -- ++(240:0.4cm) node[below, xshift=-0.5pt, inner sep=0,font=\scriptsize] {$0$};
	\draw (n4) -- ++(300:0.4cm) node[below, xshift=0.5pt, inner sep=0,font=\scriptsize] {$0$};
	\draw (n4) -- ++(320:0.6cm) -- (t3);	
	
	\node (H) at ($(n1)-(0.9, 0.3)$) {$H$};
	\node (I2) at ($(n2)-(0.9, 0.3)$) {$\mathbb{I}_2$};
\end{tikzpicture}}
\caption{Creation of $H\otimes\mathbb{I}_2$ using decision diagrams}
\label{fig:tensordd}
\end{figure}

\begin{example}\label{ex:tensordd}
    The $2\times 2$ matrix of the Hadamard gate was extended to a $4\times 4$ representation by computing the tensor product with the $2\times 2$ identity matrix~$\mathbb{I}_2$. \autoref{fig:tensordd} now illustrates this process using decision diagrams.
\end{example}

Decision diagrams do not only allow one to efficiently represent quantum states and matrices, but also to manipulate them. In the following, 
we illustrate their application for two particularly important design tasks in quantum computing: \emph{quantum circuit simulation} and \emph{quantum circuit verification}.

\subsection{Simulation of Quantum Circuits}\label{sec:simulation} 

The simulation of a quantum circuit $G$ consisting of quantum operations $g_0,\dots,g_{m-1}$ entails the consecutive calculation of the matrix-vector product between the state vector $\ket{\varphi}$ and the current operation matrix~$U_i$ (corresponding to gate $g_i$) until all operations have been applied.
The following example sketches how matrix-vector multiplication is realized on decision diagrams.  

\begin{figure*}[t]
	\centering
	\resizebox{0.99\linewidth}{!}{
		\begin{tikzpicture}[terminal/.append style={draw,rectangle,inner sep=2pt}]
			\matrix[ampersand replacement=\&,every node/.style={vertex},column sep={0.8cm,between origins},row sep={1cm,between origins}] (qmdd) {
				\& \& \node (top)[draw = none] {};\& \& \\
				\& \& \node (n1) {$q_{i}$};\& \& \\
				\node[dashed, xshift=0.4cm] (n2) {$\phantom{q_i}$}; \& \node[dashed, xshift=0.4cm] (n3) {$\phantom{q_i}$}; \& \node[dashed, xshift=0.4cm] (n4) {$\phantom{q_i}$}; \& \node[dashed, xshift=0.4cm] (n5) {$\phantom{q_i}$}; \& \\
			};
			
			\draw (top) -- (n1);
			
			\draw (n1) -- ++(240:0.6cm) -- (n2);
			\draw (n1) -- ++(260:0.6cm) -- (n3);
			\draw (n1) -- ++(280:0.6cm) -- (n4);
			\draw (n1) -- ++(300:0.6cm) -- (n5);
			
			\matrix[right=0.25cm of qmdd, ampersand replacement=\&,every node/.style={vertex},column sep={0.8cm,between origins},row sep={1cm,between origins}] (qmdd2) {
				\& \node (top2)[draw = none] {};\& \\
				\& \node (m1) {$q_{i}$};\& \\
				\node[dashed, xshift=0.4cm] (m2) {$\phantom{q_i}$}; \& \node[dashed, xshift=0.4cm] (m3) {$\phantom{q_i}$}; \& \\
			};
			
			\draw (top2) -- (m1);
			
			\draw (m1) -- ++(240:0.6cm) -- (m2);
			\draw (m1) -- ++(300:0.6cm) -- (m3);				
			
			\matrix[right=0.5cm of qmdd2, ampersand replacement=\&,every node/.style={vertex},column sep={1.5cm,between origins},row sep={1cm,between origins}] (qmdd3) {
				\& \node (top3)[draw = none] {};\& \\
				\& \node (o1) {$q_{i}$};\& \\
				\node[dashed, xshift=0.75cm] (o2) {$\phantom{q_i}$}; \& \node[dashed, xshift=0.75cm] (o3) {$\phantom{q_i}$}; \& \\
			};
			
			\draw (top3) -- (o1);
			
			\draw (o1) -- ++(240:0.6cm) -- (o2);
			\draw (o1) -- ++(300:0.6cm) -- (o3);
			
			\matrix[right=-0.25cm of qmdd3, ampersand replacement=\&,every node/.style={vertex},column sep={1.5cm,between origins},row sep={1cm,between origins}] (qmdd4) {
				\& \node (top4)[draw = none] {};\& \\
				\& \node (l1) {$q_{i}$};\& \\
				\node[dashed, xshift=0.75cm] (l2) {$\phantom{q_i}$}; \& \node[dashed, xshift=0.75cm] (l3) {$\phantom{q_i}$}; \& \\
			};
			
			\draw (top4) -- (l1);
			
			\draw (l1) -- ++(240:0.6cm) -- (l2);
			\draw (l1) -- ++(300:0.6cm) -- (l3);

			\matrix[right=0.5 of qmdd4, ampersand replacement=\&,every node/.style={vertex},column sep={1.5cm,between origins},row sep={1cm,between origins}] (qmdd5) {
				\& \node (top5)[draw = none] {};\& \\
				\& \node (q1) {$q_{i}$};\& \\
				\node[dashed, xshift=0.75cm] (q2) {$\phantom{q_i}$}; \& \node[dashed, xshift=0.75cm] (q3) {$\phantom{q_i}$}; \& \\
			};
			\draw (top5) -- (q1);
			
			\draw (q1) -- ++(240:0.6cm) -- (q2);
			\draw (q1) -- ++(300:0.6cm) -- (q3);

			\draw (n2.south) node[anchor=north] {$U_{00}$};
			\draw (n3.south) node[anchor=north] {$U_{01}$};
			\draw (n4.south) node[anchor=north] {$U_{10}$};
			\draw (n5.south) node[anchor=north] {$U_{11}$};
			
			\draw (m2.south) node[anchor=north] {$\alpha_{0\ldots}$};
			\draw (m3.south) node[anchor=north] {$\alpha_{1\ldots}$};
			
			\draw (o2.south) node[anchor=north] {$U_{00}\cdot \alpha_{0\ldots}$};
			\draw (o3.south) node[anchor=north] {$U_{10}\cdot \alpha_{0\ldots}$};
			
			\draw (l2.south) node[anchor=north] {$U_{01}\cdot \alpha_{1\ldots}$};
			\draw (l3.south) node[anchor=north] {$U_{11}\cdot \alpha_{1\ldots}$};
			
			\draw ($(n5)!0.5!(m2) + (0,0.5)$) node {$\times$};
			
			\draw ($(o3)!0.5!(l2) + (0,0.5)$) node {$+$};
			
			\draw ($(m3)!0.5!(o2) + (0,0.5)$) node {$=$};
			\draw ($(l3)!0.5!(q2) + (0,0.5)$) node {$=$};
			
			\draw ($(q2.south)+(0,-1.3)$) node[anchor=north, label={[align=center]$U_{00}\cdot \alpha_{0\ldots}$ \\$+$\\ $U_{01}\cdot \alpha_{1\ldots}$}] {};
			\draw ($(q3.south)+(0,-1.3)$) node[anchor=north, label={[align=center]$U_{10}\cdot \alpha_{0\ldots}$ \\$+$\\ $U_{11}\cdot \alpha_{1\ldots}$}] {};
			
	\end{tikzpicture}}
	\caption{Recursive structure of multiplication and addition using decision diagrams}
	\label{fig:qmdd_mult}
\end{figure*}
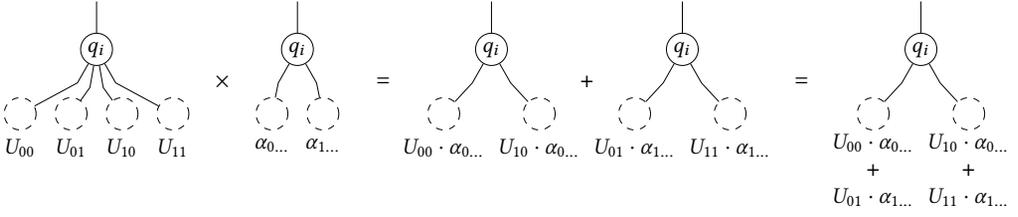

\begin{example}\label{ex:mxv}
    The multiplication of a (unitary) matrix $U$ and a (state) vector $\ket{\varphi}$ can be broken down into sub-computations as follows: 
    \begin{equation*}
        \begin{bmatrix}
        U_{00} & U_{01} \\ U_{10} & U_{11}
        \end{bmatrix}\cdot
        \begin{bmatrix}
        \alpha_{0\ldots} \\ \alpha_{1\ldots}
        \end{bmatrix} =
        \begin{bmatrix}
        (U_{00}\alpha_{0\ldots} + U_{01}\alpha_{1\ldots}) \\ (U_{10}\alpha_{0\ldots} + U_{11}\alpha_{1\ldots})
        \end{bmatrix}.
    \end{equation*}
    Now, the $U_{ij}$-submatrices precisely correspond to the four successors of the matrix's root node, while the $\alpha_{i\ldots}$-subvectors correspond to the two successors of the state vector's root node. 
    This is illustrated in \autoref{fig:qmdd_mult}.
    Thus, by recursively decomposing these sub-computations further until only operations on complex numbers remain, an efficient scheme for matrix multiplication using decision diagrams is devised\footnote{Decision diagram packages employ several implementation techniques in order to further exploit possible redundancies and to reduce the number of computations necessary, see, e.g.,~\cite{DBLP:conf/iccad/ZulehnerHW19}.}. 
\end{example}

Measuring the resulting state, i.e., sampling from the corresponding decision diagram, can be efficiently conducted by a randomized single-path traversal of the decision diagram~\cite{hillmich2020just}. At each node, the squared magnitude of the left (right) successor gives the probability of the qubit associated to the node being $\ket{0}$ ($\ket{1}$), while the probability of an individual basis state is the product of all probabilities along the path.
In contrast to quantum computations on real quantum computers, measurements of classically simulated quantum states can be conducted \mbox{non-destructively}, i.e., they can be repeated on the same state without having to repeat the whole calculation.

\subsection{Verification of Quantum Circuits}\label{sec:verification}

In order to realize a conceptual quantum algorithm on an actual device, the algorithm's description is transformed through various levels of abstraction---including steps usually called compilation, synthesis, transpilation, mapping, and/or similar. To this end, several methods have been proposed~\cite{willeMappingQuantumCircuits2019,zulehnerEfficientMethodologyMapping2019,muraliNoiseadaptiveCompilerMappings2019,matsuoReducingOverheadMapping2019,tanOptimalLayoutSynthesis2020,lao2020timing,zulehner2018compiling}.
During this process, it is of utmost importance to guarantee that the resulting circuit is still functionally equivalent to the original algorithm. 
The functionality of a quantum circuit is described by the unitary system matrix $U$ arising from the matrix-matrix multiplications of the individual gate descriptions.
Thus, the equivalence of two circuits can be reduced to the comparison of their system matrices.
In the following, we illustrate this verification scenario using the \emph{Quantum Fourier Transform}~(QFT, a popular building block in many quantum algorithms)~\cite{nielsenQuantumComputationQuantum2010} as an example.

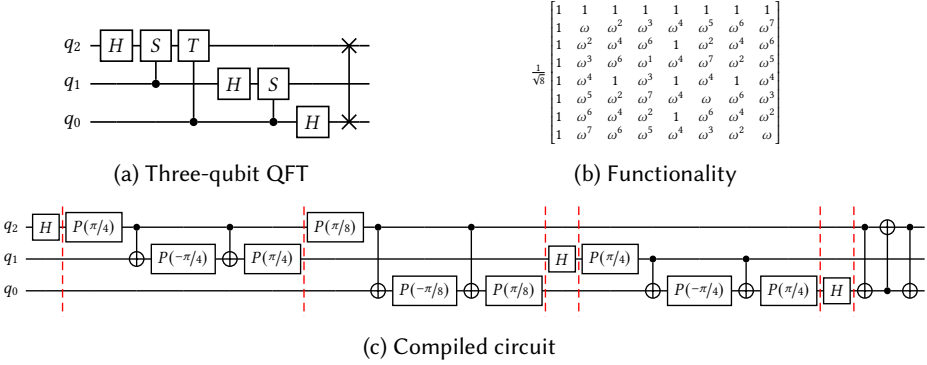
\begin{figure*}[t]
	\centering
	\begin{subfigure}[b]{0.35\linewidth}
		\centering
		\resizebox{0.9\linewidth}{!}{
			\begin{quantikz}[column sep=5pt, row sep={0.65cm,between origins}, ampersand replacement=\&]
				\lstick{$q_2$} \& \gate{H} \& \gate{S} \& \gate{T} \& \qw \& \qw \& \qw \& \swap{2} \& \qw \\
				\lstick{$q_1$} \& \qw \& \ctrl{-1} \& \qw \& \gate{H} \& \gate{S} \& \qw \& \qw \& \qw\\
				\lstick{$q_0$} \& \qw \& \qw \&  \ctrl{-2} \& \qw \& \ctrl{-1} \& \gate{H} \& \targX{}\& \qw
			\end{quantikz}
		}
		\caption{Three-qubit QFT}
		\label{fig:qftcirc}
	\end{subfigure}%
	\quad
	\begin{subfigure}[b]{0.45\linewidth}
		\centering
		\resizebox{0.55\linewidth}{!}{
			$\frac{1}{\sqrt{8}}
			\begin{bmatrix}
				1 & 1 & 1 & 1 & 1 & 1 & 1 & 1 \\
				1 & \omega & \omega^2 & \omega^3 & \omega^4 & \omega^5 & \omega^6 & \omega^7\\
				1 & \omega^2 & \omega^4 & \omega^6 & 1 & \omega^2 & \omega^4 & \omega^6\\
				1 & \omega^3 & \omega^6 & \omega^1 & \omega^4 & \omega^7 & \omega^2 & \omega^5\\
				1 & \omega^4 & 1 & \omega^3 & 1 & \omega^4 & 1 & \omega^4\\
				1 & \omega^5 & \omega^2 & \omega^7 & \omega^4 & \omega & \omega^6 & \omega^3\\
				1 & \omega^6 & \omega^4 & \omega^2 & 1 & \omega^6 & \omega^4 & \omega^2\\
				1 & \omega^7 & \omega^6 & \omega^5 & \omega^4 & \omega^3 & \omega^2 & \omega
			\end{bmatrix}$
		}
		\caption{Functionality}
		\label{fig:qftmatrix}
	\end{subfigure}%
	
	\begin{subfigure}[b]{\linewidth}
		\centering
		\resizebox{0.9\linewidth}{!}{
			\begin{quantikz}[column sep=4pt, row sep={0.65cm,between origins}, ampersand replacement=\&]
				\lstick{$q_2$} \& \gate{H}\slice{} \& \gate{P(\nicefrac{\pi}{4})} \& \ctrl{1} \& \qw \& \ctrl{1} \& \qw\slice{} \& \gate{P(\nicefrac{\pi}{8})} \& \ctrl{2} \& \qw \& \ctrl{2} \& \qw \slice{}\& \qw\slice{}\& \qw\& \qw\& \qw\& \qw\& \qw\slice{}\& \qw\slice{} \& \ctrl{2} \& \targ{} \& \ctrl{2} \& \qw \\
				\lstick{$q_1$} \& \qw \& \qw \& \targ{} \& \gate{P(\nicefrac{-\pi}{4})} \& \targ{} \& \gate{P(\nicefrac{\pi}{4})} \& \qw\& \qw\& \qw\& \qw\& \qw\& \gate{H} \& \gate{P(\nicefrac{\pi}{4})} \& \ctrl{1} \& \qw \& \ctrl{1} \& \qw\& \qw\& \qw\& \qw\& \qw\& \qw \\
				\lstick{$q_0$} \& \qw\&\qw\&\qw\&\qw\&\qw\&\qw\&\qw\&\targ{}\& \gate{P(\nicefrac{-\pi}{8})} \& \targ{} \& \gate{P(\nicefrac{\pi}{8})} \& \qw \& \qw \& \targ{} \& \gate{P(\nicefrac{-\pi}{4})} \& \targ{} \& \gate{P(\nicefrac{\pi}{4})} \& \gate{H} \& \targ{} \& \ctrl{-2} \& \targ{} \& \qw
			\end{quantikz}
		}
		\caption{Compiled circuit}
		\label{fig:qftdecomp}
	\end{subfigure}%
	\hfill
	\caption{The Quantum Fourier Transformation and its functionality}
	\label{fig:qft}
\end{figure*}

\begin{example}
The circuit of the three-qubit QFT is shown in \autoref{fig:qftcirc}. It consists of Hadamard gates, controlled phase gates, i.e., rotations with an angle that is a certain fraction of $\pi$ (e.g., $S=p(\nicefrac{\pi}{2})$, $T=p(\nicefrac{\pi}{4})$) controlled on the value of another qubit, and a SWAP gate (which swaps the value of the two qubits indicated by $\times$). The latter two types of gates are not native to any current quantum computer and, thus, need to be compiled into sequences of gates that are supported.
\autoref{fig:qftdecomp} shows one possible compiled version of the abstract QFT circuit.
Both circuits realize the functionality described by the $8\times 8$ matrix shown in \autoref{fig:qftmatrix}, where $\omega=e^{i\pi/4}=\sqrt{i} = \nicefrac{1+i}{\sqrt{2}}$.
\end{example}

While conceptually simple, the exponential size of the involved matrices quickly renders straightforward techniques based on matrices infeasible. 
Decision diagrams are a prominent candidate for checking the equivalence of two circuits since they (1) can represent quantum functionality compactly in many cases, and, (2) still offer a canonical representation (with respect to a given variable order and normalization scheme). 
Thus, the equivalence of two decision diagrams can be concluded by comparing their root pointers and the corresponding edge weights in most implementations.

\begin{figure}[tb]
	\centering
	\includegraphics[width=0.8\linewidth]{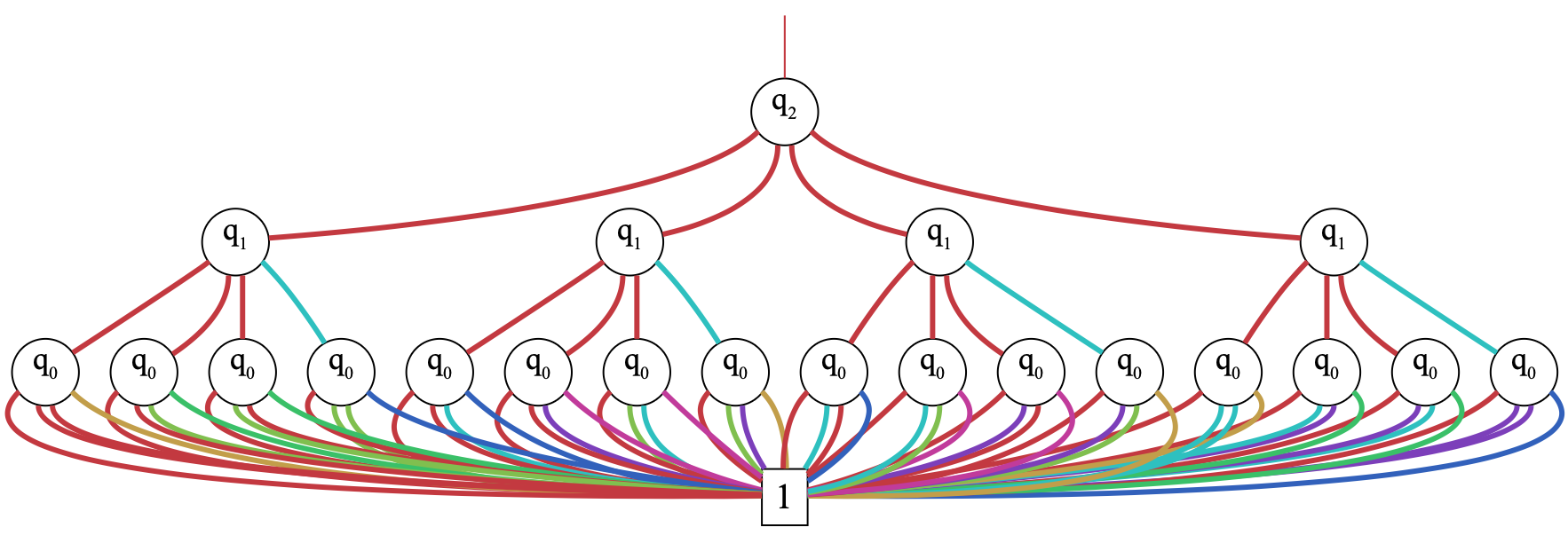}
	\caption{Decision diagram for the functionality of the three-qubit QFT} 
	\label{fig:qftdd}
\end{figure}

\begin{example}
	Constructing the decision diagrams for the two circuits shown in \autoref{fig:qftcirc} and \autoref{fig:qftdecomp}, respectively, results in a decision diagram as shown in \autoref{fig:qftdd} in both cases\footnote{For sake of readability, edge weights are not explicitly annotated to this decision diagram anymore. Instead, a color code is used (also explained in detail later in \autoref{sec:visualization} and \autoref{fig:hls}).}.
	Hence, both circuits are considered equivalent.
\end{example}

As seen in the previous example, decision diagrams can still grow exponentially large in the worst case---again presenting a severe obstacle for verifying the correctness of circuits.
However, as recently shown in~\cite{burgholzerAdvancedEquivalenceChecking2021}, this complexity can be drastically reduced in many cases by exploiting the inherent reversibility of quantum operations.
The general idea is as follows:
If two quantum circuits $G$ and $G^\prime$ are equivalent, then concatenating the first circuit~$G$ with the inverse~$G^{\prime -1}$ of the second circuit would realize the identity function~$\mathbb{I}$. The potential now lies in the order in which the operations from either circuit are applied.  
Whenever a strategy can be employed so that the respective gates from $G$ and $G^{\prime}$ are applied such that the yielded decision diagrams remain close to the identity representation, the entire procedure can be conducted efficiently with low memory usage (e.g., in case of verifying the results of compilation flows~\cite{burgholzerVerifyingResultsIBM2020}). 

\begin{example}\label{ex:gginv}
	Consider again the two circuits realizing the QFT shown earlier in \autoref{fig:qft}. Then, their equivalence can be concluded by (1) starting with a decision diagram resembling the identity, (2) applying one gate from the circuit shown in \autoref{fig:qftcirc}, and, (3) applying all gates from the circuit shown in \autoref{fig:qftdecomp} up to the next barrier (indicated by dashed lines in \autoref{fig:qftdecomp}).	
	The resulting decision diagram resembles the identity and, hence, the equivalence of both circuits can be concluded.
	Conducting the verification in this fashion only requires a maximum of 9 nodes (as opposed to $21$ nodes for building the entire system matrix).
\end{example}

\section{Visualizing Decision Diagrams in Design Automation}\label{sec:visualization}

In the previous section, we have shown that decision diagrams provide a promising basis for important design tasks such as simulation and verification. 
However, users of the corresponding tools often do not have a corresponding background or an intuition about how these methods based on decision diagrams work and what their strengths and limits are.
In an effort to make decision diagrams for quantum computing more accessible, we developed a tool which visualizes quantum decision diagrams and allows to explore their behavior when used in the design tasks covered above.
This is similar to tools such as Quirk~\cite{quirk} but with a distinct focus on decision diagrams.
To keep the effort of using the visualization as small as possible, the tool has been implemented as a web tool which can be used simply by accessing \url{https://iic.jku.at/eda/research/quantum_dd/tool}.
In the following, we illustrate how our tool (1) visualizes decision diagrams for vectors and matrices, (2) can be used for simulating quantum circuits, and (3) can be used for verifying quantum circuits.

\subsection{Styling Decision Diagrams}\label{sec:visdds}

\begin{figure}[tb]
	\centering
	\begin{subfigure}[t]{0.29\linewidth}
		\centering
		\includegraphics[width=0.38\linewidth]{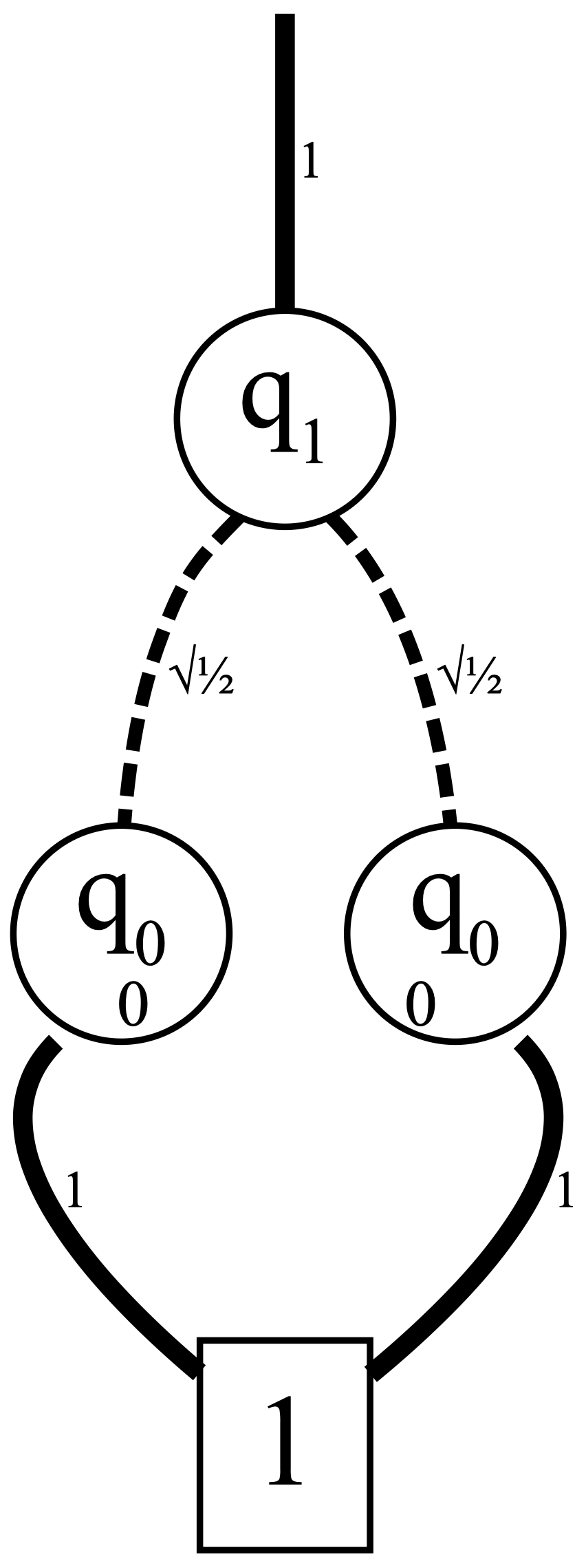}
		\caption{\enquote{Classic} mode}
		\label{fig:classic}
	\end{subfigure}%
	\hfill
	\begin{subfigure}[t]{0.39\linewidth}
		\centering
		\includegraphics[width=0.7\linewidth]{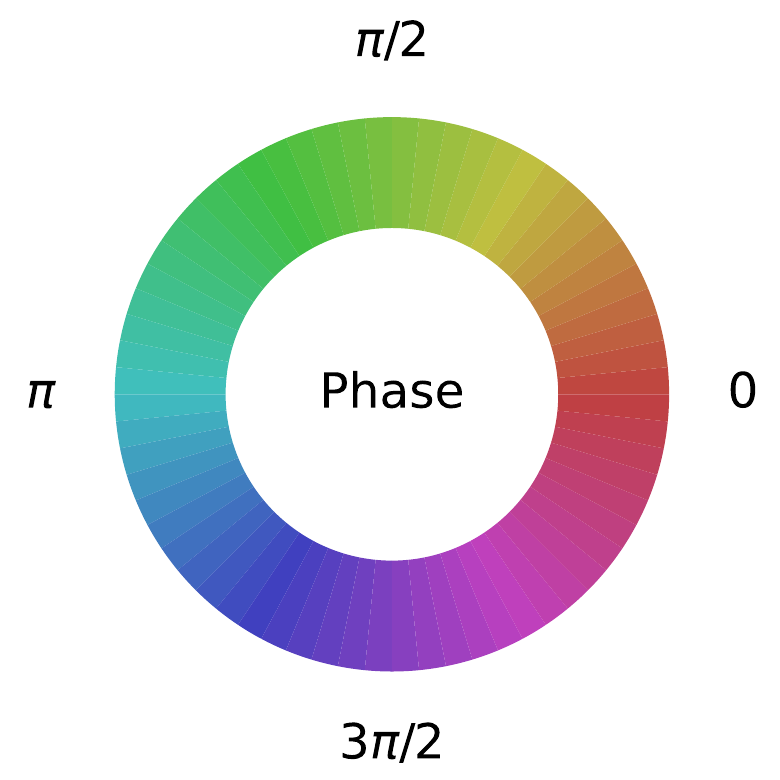}
		\caption{HLS color wheel for phase encoding}
		\label{fig:hls}
	\end{subfigure}%
	\hfill
	\begin{subfigure}[t]{0.32\linewidth}
		\centering
		\includegraphics[width=0.38\linewidth]{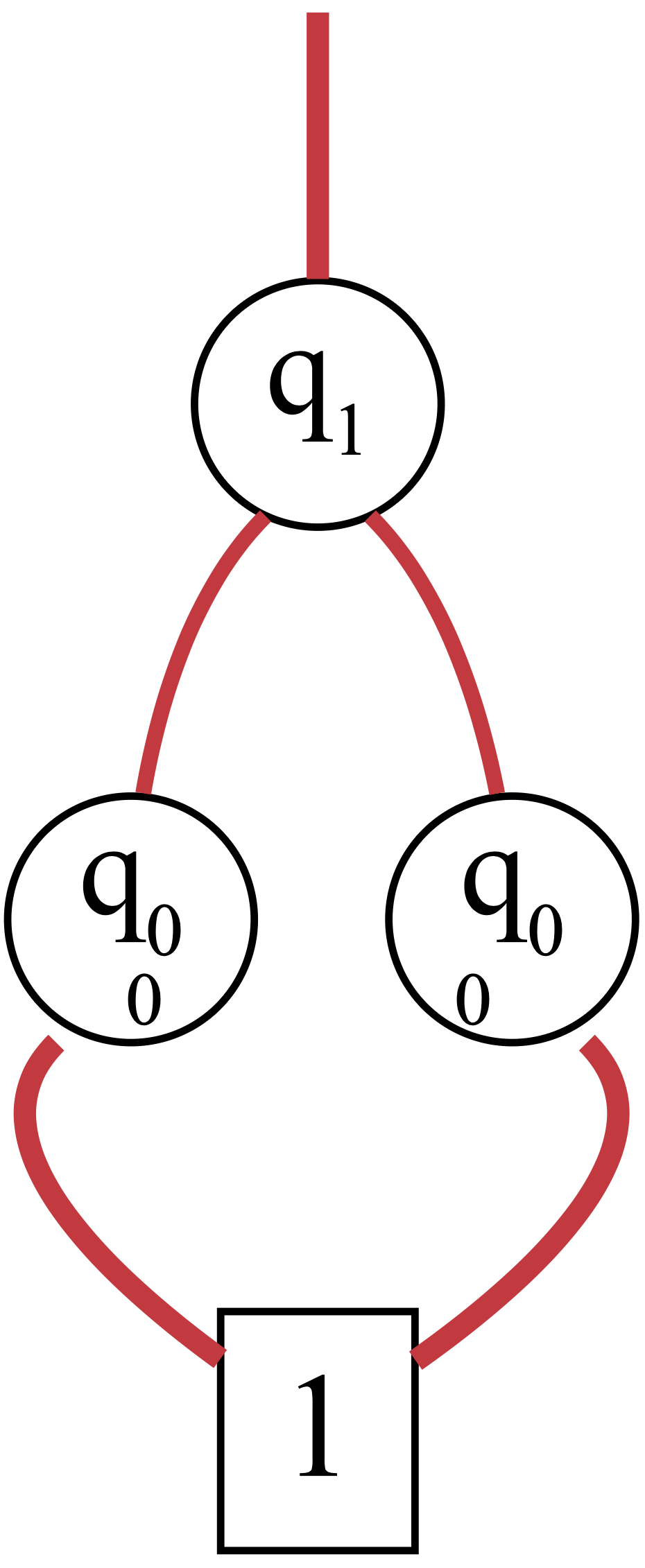}
		\caption{Colored edge weights}
		\label{fig:colored}
	\end{subfigure}
	\caption{Visualization options for vector decision diagrams}
	\label{fig:vectorvis}
\end{figure}

In order to provide the most accessible user interface possible, the tool offers several options for customizing how decision diagrams are visualized.
\autoref{fig:vectorvis} illustrates the available styles for  
decision diagrams representing vectors, i.e.,~states of a quantum system.
The \enquote{classic} mode (see \autoref{fig:classic}) offers  a look and feel that is most similar to what is found in research papers (as, e.g., shown throughout \autoref{sec:ddsapp}).
Edges with a corresponding weight not equal to $1$ are drawn using dashed lines and $0$-stubs are retracted into the nodes themselves.
Since the explicit annotation of edge weights quickly requires lots of space and leads to unreadable decision diagrams, there is also an option for removing these edge labels. Instead the magnitude of an edge weight can be reflected by the thickness of the line, while its complex phase can be color-coded using the HLS color wheel shown in \autoref{fig:hls}. Examples using this color code are shown in \autoref{fig:qftdd} and \autoref{fig:colored}. 
In addition to the above, the tool provides a more \enquote{modern} look for the decision diagram nodes, where the connection to the underlying state vector is expressed in a more straightforward fashion (shown later in \autoref{fig:simulation}).
Such a \enquote{modern} look is also available for matrices (shown later in \autoref{fig:ver}).
These should allow less accommodated users to more easily grasp the concept of decision diagrams.
In the following, we provide a deeper look into the individual features of the tool and how simulation and verification of quantum circuits can be conducted using the tool.

\subsection{Simulation of Quantum Circuits}\label{sec:vissim}

The simulation feature of the tool provides a settings panel, an algorithm box for entering or loading a quantum algorithm/circuit, 
and an interactive decision diagram box which displays the current system state in terms of a decision diagram, as well as the state vector currently represented by the decision diagram.
Loading quantum algorithms/circuits into the tool is as easy as drag- and dropping an algorithm/circuit file (in either \emph{.qasm}~\cite{cross2017open} or \emph{.real}~\cite{WGT+:2008} format) into the corresponding algorithm box, or starting to enter your own description using one of the templates provided in the \enquote{Example Algorithms} list.
Once a valid algorithm/circuit has been entered/loaded in the algorithm box, the simulation can be controlled using the navigation buttons below it: 
\begin{itemize}
	\item {\tiny$\rightarrow/\leftarrow$} : Go one step forward or backward. Can be used to step through the simulation.
	\item {\tiny$\twoheadrightarrow/\twoheadleftarrow$} : Go straight to the end (or the next special operation; see below) or back to the beginning.
	\item {\tiny$\triangleright/\vert\vert$} : Start/Pause a slide show where the simulation advances step-by-step in an automated fashion.
\end{itemize}
Some operations are considered \emph{special operations} since they do not directly correspond to the application of a unitary matrix:

\begin{figure}[t]
	\centering
	\begin{subfigure}[b]{0.45\linewidth}
		\centering
		\includegraphics[width=0.9\linewidth]{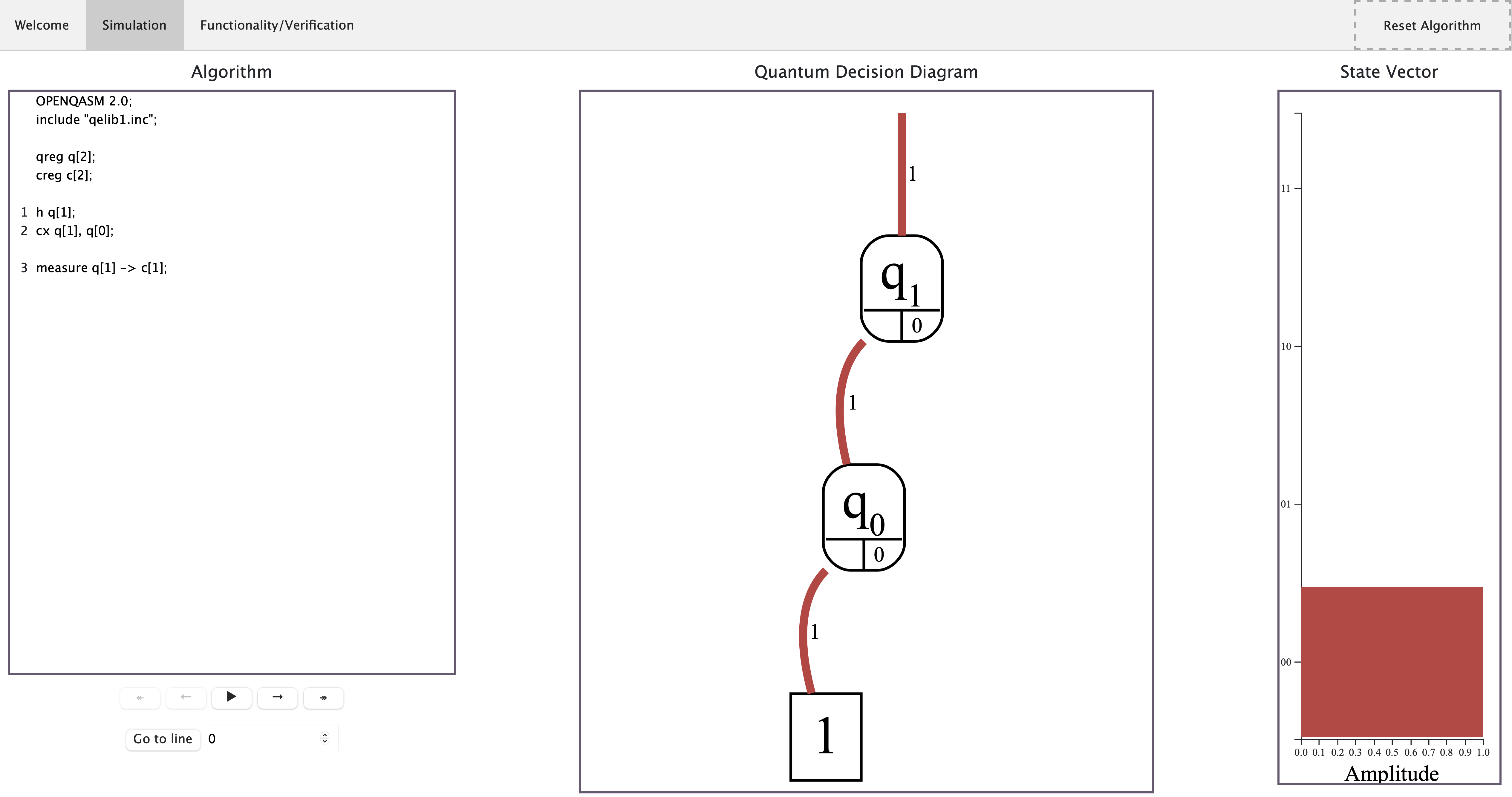}
		\caption{Initial state $\ket{00}$}
		\label{fig:sim_initial}
	\end{subfigure}%
	\hfill
	\begin{subfigure}[b]{0.45\linewidth}
		\centering
		\includegraphics[width=0.9\linewidth]{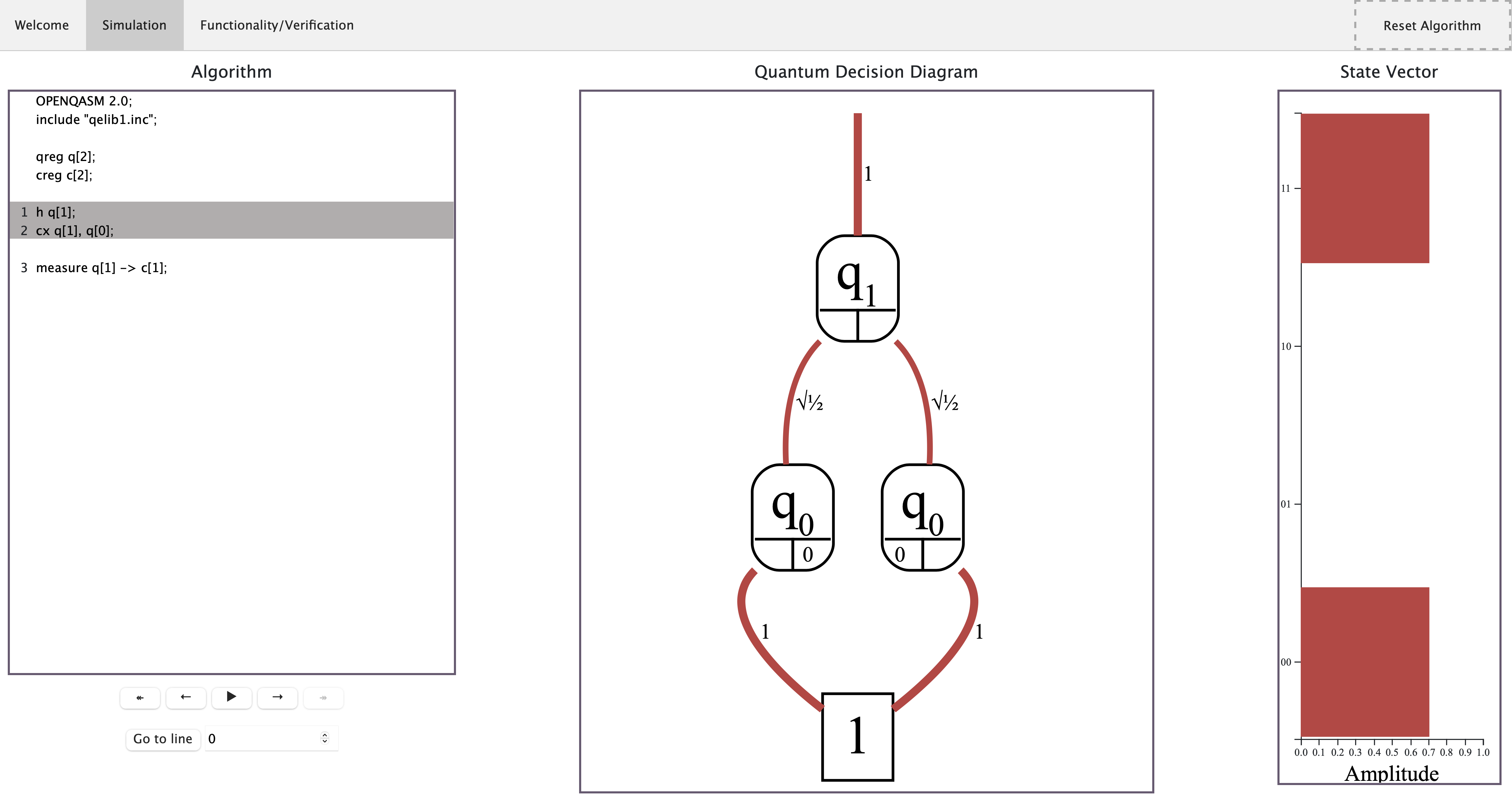}
		\caption{Resulting state $\nicefrac{1}{\sqrt{2}} \ket{00} + \nicefrac{1}{\sqrt{2}} \ket{11}$}
		\label{fig:sim_final}
	\end{subfigure}%
	
	\vspace{\floatsep}
	
	\begin{subfigure}[b]{0.45\linewidth}
		\centering
		\includegraphics[width=0.9\linewidth]{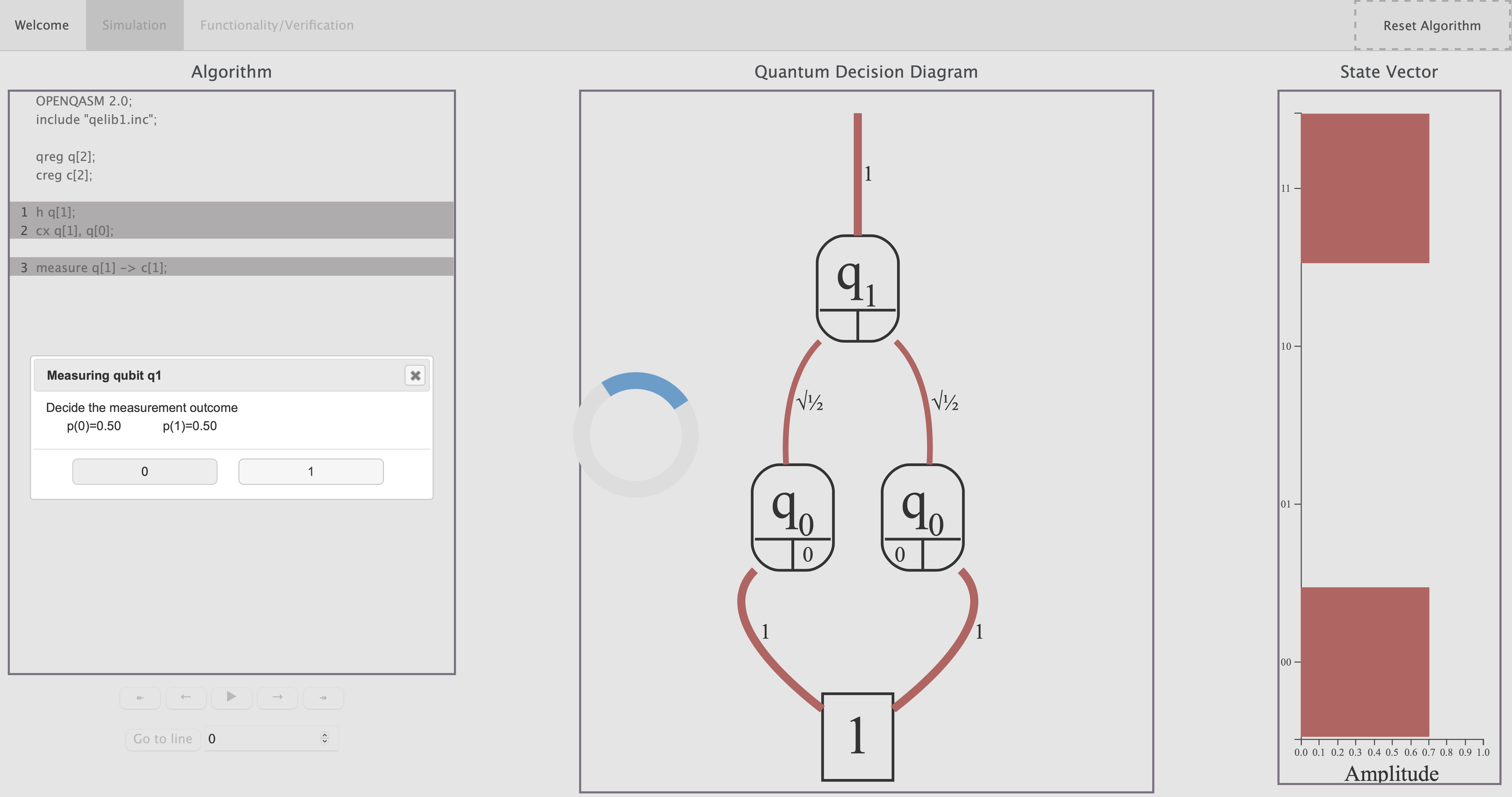}
		\caption{Measuring qubit $q_0$}
		\label{fig:sim_measure}
	\end{subfigure}%
	\hfill
	\begin{subfigure}[b]{0.45\linewidth}
		\centering
		\includegraphics[width=0.9\linewidth]{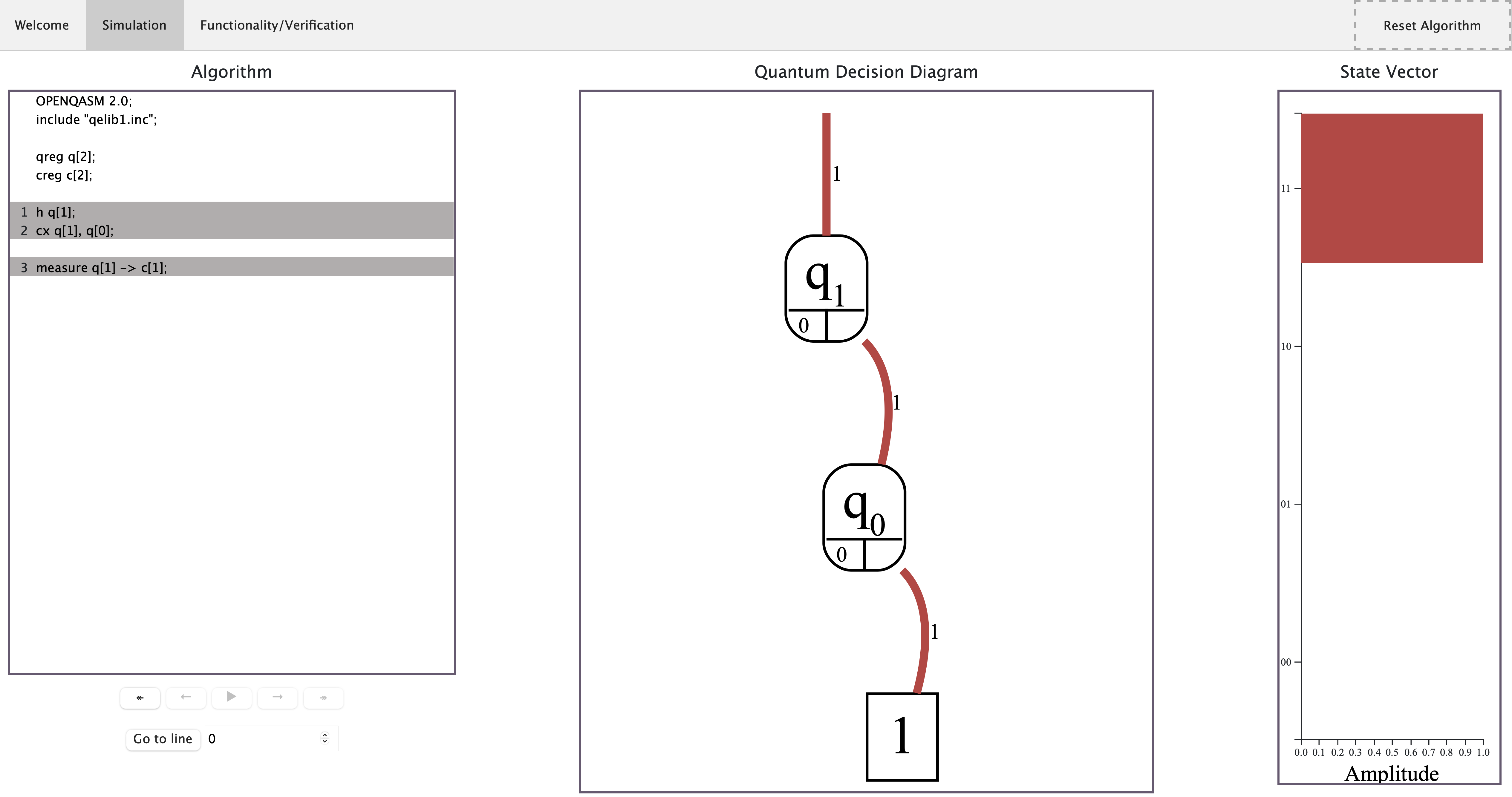}	
		\caption{Post-measurement state $\ket{11}$}
		\label{fig:sim_post}
	\end{subfigure}
	\caption{Visualizing the simulation of a circuit creating and measuring a Bell state}
	\label{fig:simulation}
\end{figure}

\begin{itemize}
	\item Barrier statements (e.g., \textit{\enquote{barrier q;}}) can be used as breakpoints when stepping forward with $\twoheadrightarrow$.
	\item Measurement operations (e.g., \textit{\enquote{measure q[0] -> c[0];}}) collapse the state of a qubit to one of its basis states. Whenever a qubit is about to get measured and it has a non-zero probability of being in either $\ket{0}$ or $\ket{1}$ (i.e., it is in superposition), a pop-up dialog appears which displays the probabilities for obtaining~$\ket{0}$ and~$\ket{1}$, respectively. Upon choosing one of the options, the decision diagram is irreversibly collapsed. Measurements also act as breakpoints due to their non-unitary (and, thus, non-reversible) nature. The tool supports OpenQASM's \mbox{classically-controlled} operations, where a certain gate is only applied if some classical bits obtained from measurements are set.
	\item Reset operations (e.g., \textit{\enquote{reset q[0];}}) discard a qubit and initialize it to $\ket{0}$ as if it were a new qubit. Mathematically, this corresponds to taking the partial trace of the whole state and, then, setting the qubit to $\ket{0}$. However, the partial trace maps pure states to mixed states and can thus in general not be represented by the same kind of decision diagram used for representing state vectors. The tool supports resets in a probabilistic fashion (similar to measurements). Whenever a reset operation is encountered where the considered qubit has a non-zero probability of being in either $\ket{0}$ or $\ket{1}$, a pop-up dialog appears which displays the probabilities for obtaining $\ket{0}$ and $\ket{1}$, respectively. Upon choosing one of the options, the other decision diagram branch is discarded and the remaining branch is set as the $\ket{0}$ branch. Resets also act as breakpoints due to their non-unitary (and, thus, non-reversible) nature.
\end{itemize}

\begin{example}
	\autoref{fig:simulation} illustrates the process of simulating the quantum circuit that creates and measures a Bell state.
	This circuit consists of two qubits, a Hadamard gate, and a controlled-NOT gate---followed by a measurement on both qubits.
	The first screenshot (\autoref{fig:sim_initial}) shows the initial quantum circuit and its state $\ket{00}$.
	Then, the two gates are applied---yielding the resulting state $\nicefrac{1}{\sqrt{2}} \ket{00} + \nicefrac{1}{\sqrt{2}} \ket{11}$ (\autoref{fig:sim_final}).
	If, now, the first qubit is to be measured, there is a \SI{50}{\percent} chance of it being either $\ket{0}$ or $\ket{1}$ (\autoref{fig:sim_measure}).
	Assume, that the measurement outcome is~$\ket{1}$. Then, the value of the second qubit is completely determined due to the entanglement of both qubits---resulting in the final state $\ket{11}$ (\autoref{fig:sim_post}).
\end{example}

\subsection{Verification of Quantum Circuits}\label{sec:visver}

The verification feature of the tool provides a similar settings panel and decision diagram box as the simulation tab, but now features two algorithm boxes.
In case only one algorithm/circuit is loaded in the left (right) algorithm box, the tool can be used to build the (inverse) functionality of the corresponding circuit.

\begin{example}
	Creating the QFT circuit shown in \autoref{fig:qftcirc} in the left algorithm box and applying all the operations precisely yields the decision diagram shown in \autoref{fig:qftdd}.
\end{example}

Once a valid algorithm/circuit has been entered in each of the algorithm boxes, their equivalence can be checked by successively applying operations from both circuits (using the corresponding {\tiny$\rightarrow/\twoheadrightarrow$} controls) and checking whether the final result resembles the identity.
As for the simulation, \emph{Barrier} statements can be used as breakpoints when stepping through both algorithms/circuits.
In contrast to simulation, \emph{Measurement}, \emph{Reset}, and \emph{Classically-Controlled Operations} are currently not supported due to their non-unitary nature.

\begin{figure}[t]
\centering
\includegraphics[width=0.9\linewidth]{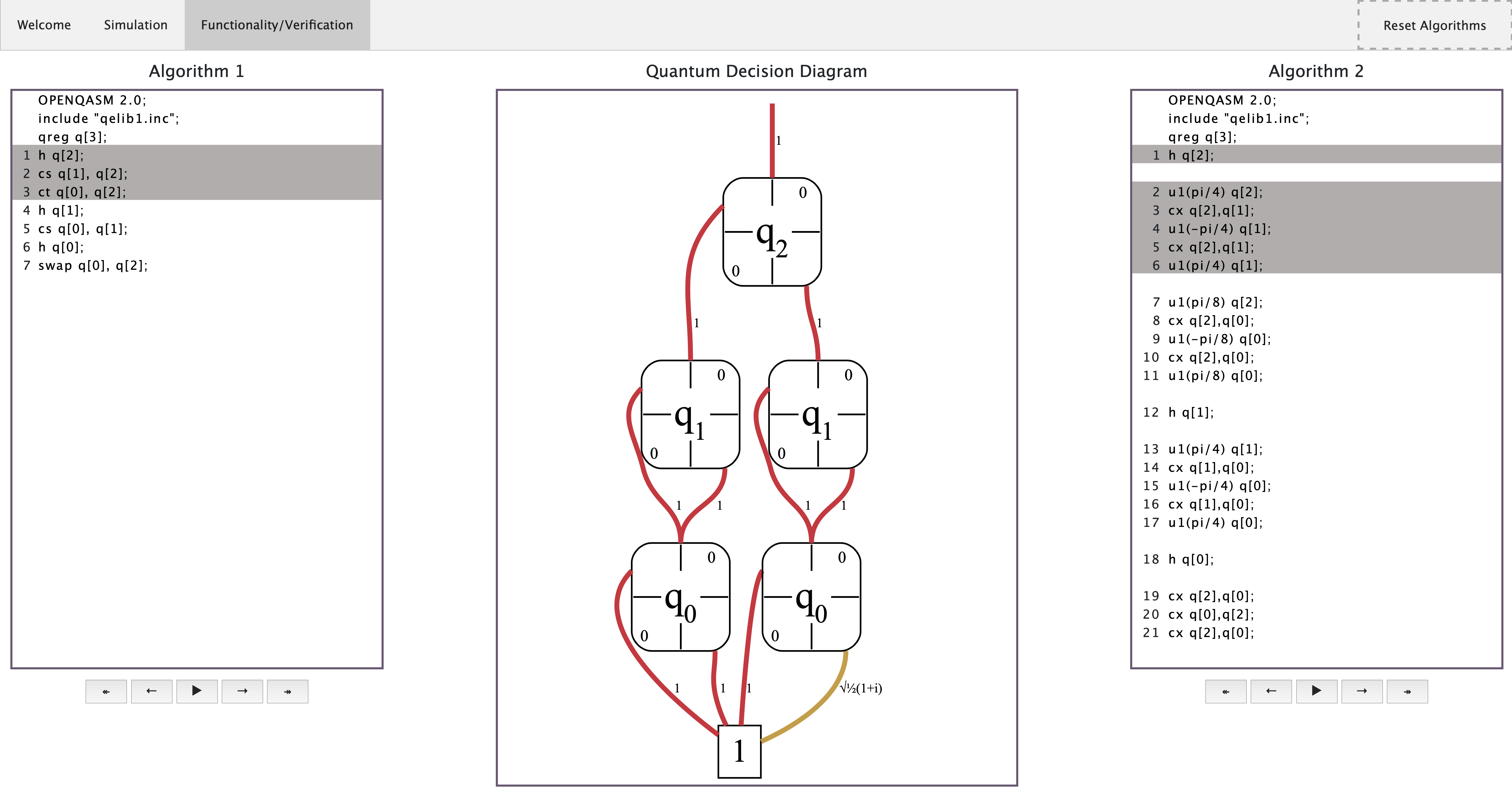}
\caption{Visualizing the verification of the QFT circuits shown in \autoref{fig:qft}}
\label{fig:ver}
\end{figure}

\begin{example}
	Consider again the two circuits realizing the three-qubit QFT shown in \autoref{fig:qftcirc} and \autoref{fig:qftdecomp}.  \autoref{fig:ver} shows how the tool is used to verify the equivalence of both circuits. The first three gates have already been applied from the left circuit, while six operations have been applied from the right circuit. 
	The corresponding decision diagram in the middle only slightly differs from the identity (by the right node labeled $q_0$).
	Continuing the computation as discussed in \autoref{ex:gginv} eventually allows to verify the equivalence of both circuits while remain close to the identity throughout the whole process.
\end{example}

\section{Decision Diagrams Within the JKQ Toolset}\label{sec:user}

The decision diagram-based methods discussed in \autoref{sec:simulation} and \autoref{sec:verification}, respectively, have been implemented as part of the open-source JKQ quantum toolset~\cite{willeJKQJKUTools2020}.
JKQ tools offer an interface for users to leverage the power of design automation for quantum computing as a black box: The user provides the input and does not have to develop a deeper understanding of the methods.
In the following, we show how the JKQ tools can be applied to two of the fundamental design problems in quantum computing: simulation (\autoref{subsec:simulation}) as well as verification (\autoref{subsec:verification}) of quantum circuits.

\subsection{Simulation of Quantum Circuits}
\label{subsec:simulation}

\emph{JKQ} offers a decision diagram-based quantum circuit simulator called \emph{DDSIM}. 
A recent case~study~\cite{DBLP:conf/ismvl/GrurlFHBW20} showed that decision diagram-based quantum circuit simulation outperforms array-based approaches whenever the (intermediate) quantum states have redundancy that can be exploited for a compact representation as decision diagram. 
DDSIM is able to simulate quantum circuits defined in \emph{.qasm}~\cite{cross2017open}, \emph{GRCS}~\cite{Boixo_2018}, and \emph{.real}~\cite{WGT+:2008} format alongside parameterized instances of Grover's algorithm~\cite{Gro:96}, Shor's algorithm~\cite{DBLP:journals/siamcomp/Shor97}, and the Quantum Fourier Transformation~\cite{nielsenQuantumComputationQuantum2010}.
Using the standalone executable for JKQ DDSIM requires cloning the GitHub repository \url{https://github.com/iic-jku/ddsim} and building the tool as in the following listing\footnote{The building process requires a C++17-compatible compiler, CMake \mbox{version $\geq 3.14$}, and the boost ({\texttt{program\_options}}) library.}.
Alternatively, a pre-compiled Python package can be installed from PyPI with \lstinline|pip install jkq.ddsim|.

\begin{lstlisting}[language=bash, frame=single, numbers=none, columns=flexible]
$ git clone https://github.com/iic-jku/ddsim
$ cd ddsim
$ cmake -DCMAKE_BUILD_TYPE=Release -S . -B build
[...]
$ cmake --build build --config Release --target ddsim_simple
[...]
$ ./build/ddsim_simple --help
[displays available commands]
\end{lstlisting}

\begin{example}\label{ex:sim}
Simulating Grover's algorithm with a two qubit oracle using the JKQ DDSIM simulator can be conducted as follows: 

\begin{lstlisting}[language=bash, frame=single, columns=flexible, numbers=none]
$ ./ddsim_simple simulate --simulate_grover 2 --shots 1000 --ps
{
  "measurements": {
    "000": 503,
    "100": 497
  },
  "non_zero_entries": 2,
  "statistics": {
    "simulation_time": 0.125837,
    "measurement_time": 0.000180,
    "benchmark": "grover_2",
    "shots": 1000,
    "n_qubits": 3,
    "applied_gates": 16,
    "max_nodes": 8,
    "seed": 0
  }
}
\end{lstlisting}
Here, the parameters define the number of measurements to be performed on the final quantum state ({\verb|--shots 1000|}) and cause the simulator to 
print statistics ({\verb|--ps|}).
The output is formatted according to the JSON standard and, hence, machine readable for further processing.
\end{example}

Simulations of a given \emph{.qasm} or \emph{.real} file with the simulator can be started  
with the parameter {\verb|--simulate_file <filename>.<extension>|} (the respective format is derived from the extension).
The full set of parameters can be listed via {\verb|./ddsim_simple --help|}.
This includes 
\begin{itemize}
\item advanced techniques such as emulation~\cite{zulehner2019matrix}, which enable significant speedups for certain quantum algorithms, 
\item weak simulation~\cite{hillmich2020just}, which more faithfully mimics a physical quantum computer, and 
\item approximating  simulation~\cite{DBLP:conf/aspdac/ZulehnerHMW20,hillmich2021efficient}, which enables a finely controlled trade-off between accuracy and runtime.
\item noise-aware simulation~\cite{grurl2021Stoch, grurl2020considering}, which allows to classically simulate the effects of noise on the execution of a quantum circuit.
\item hybrid Schrodinger-Feynman simulation~\cite{burgholzerHybridSchrodingerFeynmanSimulation2021}, which allows to exploit parallelization to significantly speedup simulation for certain use cases.
\end{itemize}

\subsection{Verification of Quantum Circuits}\label{subsec:verification}

Compiling quantum algorithms results in different representations of the considered functionality, which significantly differ in their basis operations and structure but are still supposed to be functionally equivalent. Consequently, checking whether
the originally intended functionality is indeed maintained throughout all these different abstractions becomes increasingly relevant in order to guarantee an efficient, yet correct design flow.  
Existing solutions for equivalence checking of quantum circuits suffer from the complexity of the underlying problem---which has been proven to be \mbox{QMA-complete}~\cite{janzingNonidentityCheckQMAcomplete2005}.
Most notably, the need to represent matrices which require an exponential amount of memory with respect to the number of qubits quickly makes such approaches infeasible.
However, quantum mechanical characteristics, such as the inherent reversibility of quantum operations and the compact representation of the identity with decision diagrams, provide potential for more efficient equivalence checking of quantum circuits. 

\emph{JKQ} offers a quantum circuit equivalence checking (QCEC) tool~\cite{burgholzerQCEC} which explicitly exploits these characteristics based on the ideas outlined in~\cite{burgholzerAdvancedEquivalenceChecking2021,burgholzer2020improved,burgholzer2020power} and offers a strategy especially suited for verifying compilation results~\cite{burgholzerVerifyingResultsIBM2020}, as well as dedicated random stimuli generation schemes~\cite{burgholzerRandomStimuliGeneration2021}. Similar to our simulator, JKQ QCEC can be obtained from the GitHub repository \url{https://github.com/iic-jku/qcec} and can subsequently be used by building the \verb|qcec_app| CMake target.
However, there is an easier way for users to get started with the JKQ QCEC tool. Specifically, the tool is also provided as a Python package, which can be easily installed with \lstinline{pip install jkq.qcec} and also provides native integration with IBM Qiskit.

\begin{example}\label{ex:verify}
Verifying that a quantum circuit has been compiled correctly merely requires the following lines of Python: 
\begin{lstlisting}[language=Python, frame=single, columns=flexible]
from jkq.qcec import Configuration, Strategy, verify
from qiskit import QuantumCircuit, transpile

# create your quantum circuit
qc = <...> 

# append measurements to save output mapping of physical to logical (qu)bits
qc.measure_all() 



# compile circuit to appropriate backend using some optimization level
qc_comp = transpile(qc, backend=<...>, optimization_level=<0 | 1 | 2 | 3>) 

# set dedicated verification strategy
config = Configuration()
config.strategy = Strategy.compilationflow

# verify the compilation result
result = verify(qc, qc_comp, config)
\end{lstlisting}
\end{example}
A complete list of the available methods as well as additional configuration options can be listed via {\verb|help(qcec.Configuration)|} in Python or {\verb|qcec_app --help|} when using the commandline app.

\section{Developer's Perspective}\label{sec:developer}
The design automation tools described in the previous sections are quite powerful by themselves; nonetheless, there is always room for improvement and additional features.
Because of this, the visualization and the \emph{JKQ} tools are open source and developers are kindly invited to extend or modify the methods at their own discretion.
This section gives a brief overview of how the tools work internally and serves as a starting point for the interested developer wanting to take a deeper dive.

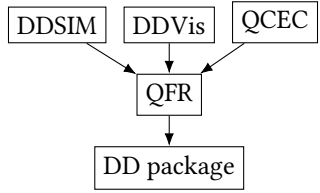
\begin{figure}[tbp]
	\centering
	\begin{tikzpicture}[every node/.style={draw, rectangle}, node distance=0.4cm and 0.4cm]
		\node (ddpackage) {DD package};
		\node[above=of ddpackage] (qfr) {QFR};
		
		\node[above left=of qfr] (ddsim) {DDSIM};
		\node[above=of qfr] (ddvis) {DDVis};
		\node[above right=of qfr] (qcec) {QCEC};		
		
		\begin{scope}[-Latex]
			\draw (ddsim) -- (qfr);
			\draw (qcec) -- (qfr);
			\draw (ddvis) -- (qfr);
			\draw (qfr) -- (ddpackage);
		\end{scope}
	\end{tikzpicture}
	\caption{Structure of the JKQ tools}
	\label{fig:jkq-tools}
\end{figure}

The dependency relations of the JKQ tools discussed here are illustrated in \autoref{fig:jkq-tools} and the source code of the individual application can be found on GitHub as individual repositories via \url{https://github.com/iic-jku}.
Each application depends on a library referred to as \emph{Quantum Functionality Representation} (QFR), which handles the input and output of files describing quantum functionalities.
Additionally selected quantum algorithms such as Grover's search and Shor's algorithm are directly integrated as classes---allowing to programatically construct the respective quantum circuits for simulation, compilation, or verification with parameters controlling, e.g., the number qubits.
The QFR itself depends on a package providing the functionality for representing and manipulating quantum states and operations via decision diagrams~\cite{DBLP:journals/tcad/ZulehnerW19,DBLP:conf/iccad/ZulehnerHW19}.

The DD package has options that are set during compile time to enable more aggressive compiler optimizations and influence the later execution in simulation and verification. These fall primarily into one of the following categories:
\begin{itemize}
	\item \emph{Cache sizes:} The underlying routines storing and operating on decision diagrams use different caches for nodes, edges, and complex numbers. 
	There is a trade-off between larger cache sizes and better data locality. 
	Developers can adjust these values according to their needs.
	\item \emph{Floating point representation:} Across all individual projects, the generally used floating point datatype by default is {\verb|double|} (i.e., 64 bit on most platforms), as defined through the {\verb|fp|} alias (in {\verb|include/dd/Definitions.hpp|}).
	Depending on the required precision, the developer may change this to {\verb|float|} (less precision with faster execution) or {\verb|long double|} (higher precision but slower execution).
	If the precision is changed, this should be reflected in the {\verb|TOLERANCE|} (defined in {\verb|include/dd/ComplexTable.hpp|}) which mitigates effects caused by the fundamentally limited precision in the representation of complex numbers~\cite{DBLP:conf/iccad/ZulehnerHW19}.
\end{itemize}

Support for additional \enquote{hardcoded} algorithms or file formats should be integrated into the QFR, so the tools for simulation, verification, and visualization can access these new features. We recently added support for importing IBM Qiskit \verb|QuantumCircuit| objects from Python to the QFR library---allowing to integrate our tools with Qiskit through Python bindings. In the future, this could be extended to integrate with other popular quantum software frameworks.

Internally, the simulator can be used as follows to simulate the  \emph{.qasm} file \verb|shor_115_2.qasm| and conduct \num{1000} measurements:
\begin{lstlisting}[language=c++, frame=single, columns=flexible]
QuantumComputation qc1("shor_115_2.qasm");
QFRSimulator sim(qc1);
sim.Simulate();
auto samples = sim.MeasureAllNonCollapsing(1000);
\end{lstlisting}
This reads a file {\verb|shor_115_2.qasm|} into the QFR and uses the resulting  {\verb|QuantumComputation|} object to construct the simulator instance.
The actual simulation process is handled in the {\verb|Simulate|} method, where developers can start optimizing for their specific problem by creating their own simulator sub-class.
For the simulation of files each instruction in the input file is translated into the corresponding decision diagram (for unitary operations) and applied to the quantum state. \mbox{Non-unitary} operations such as measurements require separate handling in the program. Other paradigms such as \emph{approximating simulation}~\cite{DBLP:conf/aspdac/ZulehnerHMW20,hillmich2021efficient}, \emph{weak simulation}~\cite{hillmich2020just}, noise-aware simulation~\cite{grurl2021Stoch}, and hybrid Schrodinger-Feynman simulation~\cite{burgholzerHybridSchrodingerFeynmanSimulation2021} are also supported and easy to extend.

The equivalence checking methodology described in~\cite{burgholzerVerifyingResultsIBM2020,burgholzer2020improved,burgholzer2020power,burgholzerAdvancedEquivalenceChecking2021, burgholzerRandomStimuliGeneration2021} is readily extendable and offers lots of freedom for adapting to specific scenarios.
 Developers wanting to implement their own equivalence checking strategies can get started at  {\verb|ImprovedDDEquivalenceChecker.hpp|}. 
There, the {\verb|Proportional|} strategy for example is realized in the following way: 

\clearpage

\begin{lstlisting}[language=c++, frame=single, columns=flexible]
int ratio1 = std::max(1, qc1.getNops()/qc2.getNops());
int ratio2 = std::max(1, qc2.getNops()/qc1.getNops());
while (it1 != end1 && it2 != end2) {
	for (int i=0; i<ratio1 && it1!=end1; ++i, ++it1) 
		applyGate(*it1, results.result, perm1, LEFT);
	for (int i=0; i<ratio2 && it2!=end2; ++i, ++it2) 
		applyGate(*it2, results.result, perm2, RIGHT);
}
\end{lstlisting}
In {\verb|CompilationFlowEquivalenceChecker.hpp|}, the dedicated compilation flow verification strategy can easily be extended to anticipate further optimizations, or adapt to compilation flows different than IBM Qiskit~\cite{QiskitNoDOI}.
Currently, QCEC supports different notions of equivalence, i.e., \enquote{unitary equivalence}, \enquote{equivalence up to global phase}, and \enquote{equivalence of measurement outcomes}. In the future, the methodology could also be extend to support further notions of equivalence.
This includes \enquote{equivalence up to relative phase}, \enquote{equivalence up to permutation of the outputs} or---even more general---\enquote{equivalence up to relabeling of the qubits}.

The JKQ decision diagram visualization tool DDVis is designed as a NodeJS application that builds on top of our QFR library. It uses the \verb|dd::export2Dot()| functionality of the DD package to generate a GraphViz~\cite{Gansner00anopen} representation of the considered decision diagrams, which is, in turn, rendered using \verb|d3-graphviz|. At the moment, it provides basic support for visualizing the simulation and verification of quantum circuits. The simulation part might be extended in the future to accommodate the approximation capabilities of the JKQ~DDSIM simulator. Similarly, the verification part might be extended to automatically conduct one of the available verification schemes from the JKQ~QCEC tool and provide an indicator whether the two circuits are considered equivalent at any particular time. 

For more details on the implementations, we refer to the documentations in the respective repositories at \url{https://github.com/iic-jku/} which include links to pre-compiled packages on the PyPI service.

\section{Conclusions}\label{sec:conclusions}

In an effort to bridge the gap between the design automation and the quantum community and to make decision diagrams for quantum computing more accessible, we provided a summary on how decision diagram techniques can be used for the design of quantum circuits.
Additionally, we presented an easily-accessible tool which visualizes quantum decision diagrams as well as their applications
and covered the usage of the \emph{JKQ} tools from two perspectives: First, for users, who want to solve their problems, but not necessarily develop a deeper understanding of the tools and how they work.
Second, for developers, who want to enhance or integrate the tools to tackle their specific problems in quantum computing.
More information about the corresponding tools is available at \url{https://github.com/iic-jku/}.
We sincerely hope that our efforts help interested researchers to learn and adopt these techniques.

\bibliographystyle{ACM-Reference-Format}
\bibliography{lit_header.bib, references.bib, library.bib}

\end{document}